\documentclass[preprint,aps,12pt,preprintnumbers,nofootinbib] {revtex4}

\pdfoutput=1

\usepackage{graphicx}
\usepackage{amssymb}
\usepackage{amsmath}
\usepackage{epstopdf}
\usepackage{subfigure}
\usepackage{setspace}

\def\dd{d\!\!{}^-\!}
\def\del{\delta\!\!\!{}^-\!}
\def\Ab{\bar{A}}
\def\wb{\bar{w}}
\def\Xb{\bar{X}}
\def\Fb{\bar{F}}
\def\etab{\bar{\eta}}
\def\hb{\bar{h}}
\def\pwwb{\partial_w\partial_{\bar{w}}}
\def\pwu{\partial_w\partial_u}
\def\pwbu{\partial_{\bar{w}}\partial_u}

\begin{document}

\title{The Kinematic Algebra From the Self-Dual Sector}

\author{Ricardo Monteiro}
\author{Donal O'Connell}
\affiliation{The Niels Bohr International Academy and Discovery Center, The Niels Bohr Institute, Blegdamsvej 17, DK-2100 Copenhagen, Denmark \vspace{0.5cm}}

\date{\today}

\begin{abstract}

We identify a diffeomorphism Lie algebra in the self-dual sector of Yang-Mills theory, and show that it determines the kinematic numerators of tree-level MHV amplitudes in the full theory. These amplitudes can be computed off-shell from Feynman diagrams with only cubic vertices, which are dressed with the structure constants of both the Yang-Mills colour algebra and the diffeomorphism algebra. Therefore, the latter algebra is the dual of the colour algebra, in the sense suggested by the work of Bern, Carrasco and Johansson. We further study perturbative gravity, both in the self-dual and in the MHV sectors, finding that the kinematic numerators of the theory are the BCJ squares of the Yang-Mills numerators. 

\end{abstract}

\maketitle

\section{Introduction}

At first glance, Yang-Mills (YM) theory and gravity seem to have little in common. This is especially true from the point of view of calculating perturbative scattering amplitudes: then the Einstein-Hilbert Lagrangian contains an infinite number of vertices, but the Yang-Mills Lagrangian contains only three and four-point vertices. While it is often said that both theories are gauge theories, the enormous differences between the theories from the point of view of computation makes this point of comparison seem rather poetic. 

But a deeper study reveals concrete relationships between gauge and gravity scattering amplitudes. Early in the first heyday of string theory, Kawai, Lewellen and Tye (KLT) derived a relation, valid in any number of dimensions, expressing any closed string tree amplitude in terms of a sum of products of two open string tree amplitudes~\cite{Kawai:1985xq}. Taking the field theory limit of the string amplitudes, these KLT relations imply in particular that any graviton scattering amplitude can be expressed in terms of a sum of products of colour-ordered gluon scattering amplitudes. The KLT relations are very simple for low point amplitudes; for example, the three-point graviton scattering amplitude is simply the square of the three-point Yang-Mills amplitude.\footnote{These three-point amplitudes are non-zero if the momenta of the particles are taken to be complex valued.} For higher-point amplitudes, however, the sum in the KLT relations becomes complicated, presenting an algorithmic obstacle to computing graviton amplitudes via KLT relations.

A more insightful ``squaring" relationship between gauge theory and gravity amplitudes was conjectured recently by Bern, Carrasco and Johansson (BCJ)~\cite{Bern:2008qj}. This relationship implies a set of identities satisfied by gauge theory amplitudes, which were proven in \cite{BjerrumBohr:2009rd,Stieberger:2009hq,Feng:2010my}. BCJ conjectured that scattering amplitudes in gravity can be obtained from Yang-Mills scattering amplitudes, expressed in a suitable form, by a remarkably simple squaring procedure. In brief, the BCJ procedure involves expressing Yang-Mills amplitudes in terms of Feynman-like diagrams which only have cubic vertices. Associated with each cubic diagram is a colour factor and also a kinematic numerator. BCJ observed two remarkable things about these kinematic numerators. First, when a set of three colour factors satisfies a Jacobi identity, then the kinematic numerators can be chosen so that they also satisfy the same Jacobi identity. Second, the graviton scattering amplitude is obtained from the Yang-Mills amplitude by simply replacing the colour factors by the kinematic numerators; this ``squaring" relationship has been proven in \cite{Bern:2010yg} assuming that appropriate gauge theory numerators exist. Explicit numerators have been described in \cite{Kiermaier,BjerrumBohr:2010hn,Mafra:2011kj}. Moreover, the relations are conjectured to hold off-shell~\cite{Bern:2010ue}. The BCJ relations have been discussed extensively in recent work, see~\cite{Sondergaard:2009za,BjerrumBohr:2010zs,Tye:2010dd,Bern:2010ue,Jia:2010nz,BjerrumBohr:2010ta,Tye:2010kg,Vaman:2010ez,Chen:2010ct,BjerrumBohr:2011kc,Bern:2011ia,BjerrumBohr:2011xe} for example. The left-right factorization of the gravitational Feynman rules to all orders in perturbation theory has been recently demonstrated by Hohm \cite{Hohm:2011dz}.

It is especially intriguing that the kinematic numerators appearing in the BCJ relations satisfy Jacobi identities. Indeed, at four-point level this relation was found in the early 1980s~\cite{Zhu:1980sz,Goebel:1980es}. The fact that these Jacobi identities are satisfied at all points strongly suggests that there is a genuine infinite dimensional Lie algebra at work in the theory. Moreover, since the gravitational amplitudes are formed by replacing Yang-Mills colour factors with these kinematic numerators, this group seems to entirely determine the gravitational scattering amplitudes. It is worth emphasizing that the kinematic numerators which BCJ discuss are gauge dependent, so that the presence of this kinematic algebra seems to be manifest only in a class of gauges. In these gauges, the relationship between Yang-Mills theory and gravity is especially simple. 

In order to study this kinematic algebra in its simplest habitat, we consider the self-dual sectors of Yang-Mills theory and of gravity. The scattering amplitudes in these sectors vanish at tree level (with the exception of the three-point amplitude). However, a scattering amplitude is a rather perverse object to study in a classical field theory. It is more natural to study solutions of the field equations; indeed, tree-level scattering amplitudes can easily be obtained by taking a limit of a series expansion of an appropriate classical solution. We find that the kinematic algebra is manifest by a simple comparison of the equations governing self-dual Yang-Mills and gravity. The algebra is associated with area-preserving diffeomorphisms in a certain two-dimensional space. Our understanding of this algebra makes it completely trivial to see that the BCJ relations extend to the classical solutions of Yang-Mills theory and gravity which describe scattering of an arbitrary number of positive helicity particles, and measuring the resulting field at a point. There is an extensive literature on the hidden symmetries of self-dual gauge theory and gravity; see e.g. \cite{Popov:1996uu,Popov:1998pc,Wolf:2004hp,Popov:2006qu}.

We further study the consequences of these results for the full Yang-Mills and gravitational theories. It is useful to consider these theories in the light-cone gauge, where they can be formulated in terms of positive and negative helicity particles. Using a procedure inspired by Chalmers and Siegel \cite{Chalmers:1998jb}, we are able to relate the MHV amplitudes to the self-dual sector, and to show that the corresponding BCJ relations are a result of the self-dual kinematic algebra.

The structure of the paper is as follows. In section~\ref{sec:bfReview} we open with a review of some background material. With this review in hand, we turn in sections~\ref{sec:sdym} and~\ref{sec:sdgr} to a discussion of the self-dual Yang-Mills and gravity theories, respectively. In these sections, we discuss the details of the kinematic algebra and BCJ relations for classical backgrounds. We discuss the implications of this algebra for the BCJ structure of MHV amplitudes of the full Yang-Mills theory in section~\ref{sec:full}, and of gravity in section~\ref{sec:gravity}. We close with some final remarks in section~\ref{sec:concl}.


\section{Background material}
\label{sec:bfReview}

Since this article deals with classical background fields applied to topics of interest in the recent literature on scattering amplitudes, it may be helpful to provide some review of our methods. We begin with a discussion of the BCJ relations, then provide an overview of the relationship between classical field solutions and tree amplitudes, and finally discuss the spinor-helicity method and a slight extension of this method which has been helpful in previous work on self-dual Yang-Mills theory and gravity.

\subsection{The BCJ relations}

Let us set the scene with a description of the BCJ relations~\cite{Bern:2008qj}. The starting point is to express the $M$-point Yang-Mills amplitudes as a sum over the cubic diagrams,
\begin{equation}
\mathcal{A}_M = g^{M-2} \sum_j \frac{c_j n_j}{D_j},
\end{equation}
where $c_j$ are the colour factors, the denominators $D_j$ are products of the Feynman propagator denominators appearing in the cubic graphs, and $n_j$ are kinematic numerators.
The cubic diagrams can be thought of as diagrams showing how the Yang-Mills structure constants are sewn together to make the various appropriate colour factors for the scattering amplitudes; at four points there are three diagrams (for the $s, t$ and $u$ channels), at five points there are 15 diagrams  and so on. Contact terms which appear in the Feynman diagram expansion of the amplitudes are assigned to appropriate cubic diagrams based on their colour structure; missing propagators in these contact terms are simply reintroduced by multiplying and dividing the contact term by such propagators.

Given this decomposition, BCJ asserted that one can always choose numerators $n_j$ such that whenever the colour factors $c_{j_1}, c_{j_2}$, and $c_{j_3}$ of three diagrams $j_1, j_2$ and $j_3$ are related by a Jacobi identity, then so are their kinematic numerators:
\begin{equation}
c_{j_1} \pm c_{j_2} \pm c_{j_3} = 0 \Rightarrow n_{j_1} \pm n_{j_2} \pm n_{j_3} = 0,
\end{equation}
where the appropriate signs depend on the definitions of the colour factors. Furthermore, the kinematic numerators can be chosen so that they enjoy the same antisymmetry properties as the colour factors under exchanging particle labels:
\begin{equation}
c_j \rightarrow -c_j \Rightarrow n_j \rightarrow - n_j.
\end{equation}
Taken together, these requirements indicate that the kinematic numerators $n_j$ have the same algebraic properties as the colour factors $c_j$. 

Given an expression for the Yang-Mills scattering amplitude in terms of numerators which satisfy the kinematic Jacobi identity, BCJ further noticed that the gravitational scattering amplitude is simply given by replacing the Yang-Mills colour factors by another copy of the kinematic numerator\footnote{More general gravity theories can also be obtained using numerator factors from two different gauge theories~\cite{Bern:2010ue}.},
\begin{equation}
\label{BCJgrav}
\mathcal{M}_M = \kappa^{M-2} \sum_j \frac{n_j n_j}{D_j} ,
\end{equation}
where $\kappa$ is the gravitational coupling.

It is worth remarking on the nature of the choices one has to make in defining appropriate numerators. Indeed, these numerators are not unique. For example, at four points the Yang-Mills scattering amplitude is written as
\begin{equation}
\mathcal{A}_4 =g^2 \left(\frac{c_s n_s}{s} + \frac{c_t n_t}{t} + \frac{c_u n_u}{u}\right). 
\end{equation}
Since $c_s + c_t + c_u = 0$, the amplitude can also be written in terms of new numerators
\begin{equation}
n'_s = n_s + s \alpha(p_i, \epsilon_j), \quad n'_t = n_t + t \alpha(p_i, \epsilon_j), \quad n'_u = n_u + u \alpha(p_i, \epsilon_j).
\end{equation}
where $\alpha$ is an entirely arbitrary function of the momenta and polarizations. Similar redefinitions are possible at higher points. BCJ described these redefinitions as generalized gauge transformations, since ordinary gauge transformations are a familiar way of moving terms between various Feynman diagrams. In the on-shell four-point case, the numerator redefinitions are such that the kinematic Jacobi identity, $n'_s + n'_t + n'_u = 0$, is maintained. However, as BCJ discuss, this is not true at higher points, even on shell. Thus, one must choose an appropriate gauge to find numerators which satisfy the kinematic Jacobi identity.

\subsection{Background fields as generating functions}

We know turn to reminding the reader of the connection between tree-level Feynman diagrams and classical field solutions \cite{Boulware:1968zz}. In a few words, tree Feynman diagrams arise as the diagrammatic representation of terms in a perturbative series for the classical solution in the presence of a source, $J$. This can be understood from the functional approach in quantum field theory. Consider the generating functional
\begin{equation}
Z[J] = \int D \phi \; e^{i (S[\phi] + J \phi)} = e^{i W[J]} ,
\end{equation}
where the fields are symbolically represented by $\phi$. $W[J]$ is the generating functional for connected correlation functions. Amplitudes are computed from the connected correlators by the LSZ procedure. Since our focus in this work is on tree level, we take the limit $\hbar \rightarrow 0$. In this limit, the path integral is dominated by solutions $\phi_{\mathrm{cl}}[J]$ of the equations of motion. Therefore,
\begin{equation}
W[J] = S[\phi_{\mathrm{cl}}[J]] + J \phi_{\mathrm{cl}}[J] .
\end{equation}
The $n$-point connected correlators are obtained by the $n$th functional differentiation of $W[J]$ with respect to $J$. However, by the equations of motion,
\begin{equation}
\frac{\delta W}{\delta J} = \left(\frac{\delta S}{\delta  \phi_{\mathrm{cl}}}  + J \right) \frac{\delta \phi_{\mathrm{cl}}[J]}{\delta J} + \phi_{\mathrm{cl}}[J] =  \phi_{\mathrm{cl}}[J] .
\end{equation}
Thus, the $n$-point connected correlators are generated by $n-1$ differentiations of the background field created by the source.

Much of the recent progress in our understanding of scattering amplitudes for massless particles has relied on the fact that the external legs of the amplitudes are on-shell, that is, their momenta $p$ satisfy $p^2 = 0$. There is a limit of these generating functionals for which all but one of the legs are on-shell. This technique has been used, for example, in the classic paper of Berends and Giele on recursion relations in gauge theory~\cite{Berends:1987me}, as well as in a beautiful paper of Brown~\cite{Brown:1992ay} to sum tree graphs contributing to threshold particle production. Let us now illustrate how this technique works in the simple setting of scalar field theory.

\begin{figure}
\centering
\includegraphics[scale=0.5]{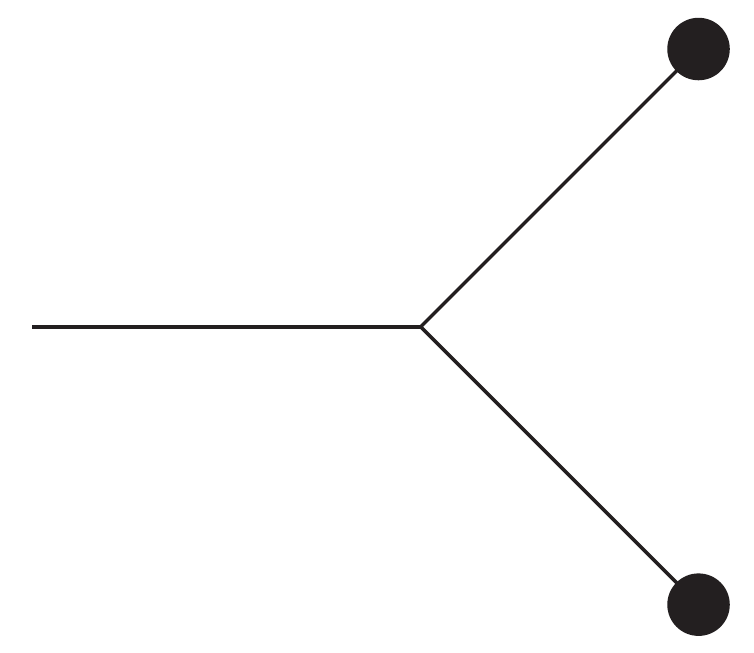}
\put(0,0){$J$}
\put(-45,15){$\phi^{(1)}$}
\put(0,30){$J$}
\caption{The first order correction $\phi^{(1)}(k)$ consists of a single interaction between two particles sourced by $J$.}
\label{fig:scalar3}
\end{figure}

We will consider a massless cubic scalar field theory, with Lagrangian (including the source)
\begin{equation}
\mathcal{L} = \frac 12 (\partial \phi)^2 - \frac{1}{3!} g \phi^3 + J \phi .
\end{equation}
The classical equation of motion is simply
\begin{equation}
\partial^2 \phi + \frac 12 g \phi^2 - J = 0 .
\end{equation}
It is convenient to Fourier transform to momentum space; then the classical equation is written as
\begin{equation}
k^2  \phi(k) - \frac 12 g \int \dd p_1 \dd p_2 \, \del (p_1 + p_2-k) \, \phi(p_1) \phi(p_2) = - J(k) ,
\end{equation}
where we have introduced a short-hand notation
\begin{equation}
\dd p \equiv \frac{d^4 p}{(2 \pi)^4} , \quad \del (p) \equiv (2 \pi)^4 \delta^4  (p) .
\end{equation}
We will solve this equation order-by-order in perturbation theory. Thus, we write
\begin{equation}
\phi(k) = \sum_{n = 0}^\infty \phi^{(n)}(k) ,
\end{equation}
where $\phi^{(n)}(k)$ is of order $g^n$. At zeroth order, we must simply solve the free equation, so we learn that
\begin{equation}
\phi^{(0)} (k)= -\frac{J(k)}{k^2} .
\end{equation}
At next to leading order, counting powers of $g$, we find that $\phi^{(1)}$ satisfies the equation
\begin{equation}
k^2 \phi^{(1)}(k) = \frac 12 \, g \int  \dd p_1 \dd p_2 \, \del (p_1 + p_2 - k ) \, \phi^{(0)}(p_1) \phi^{(0)}(p_2) ,
\end{equation}
so that 
\begin{equation}
\phi^{(1)}(k) = \frac 12 \int  \dd p_1 \dd p_2 \, \del (p_1 + p_2 - k ) \, \frac{1}{k^2}\,g\, \frac{J(p_1)}{p_1^2} \frac{J(p_2)}{p_2^2}.
\end{equation}
Diagrammatically, this equation is shown in Fig.~\ref{fig:scalar3}. One can construct the three-point scattering amplitude from this expression by functionally differentiating twice with respect to the source $J$, and amputating the three lines. That is,
\begin{equation}
\mathcal{A}_3(p_1, p_2, p_3) =-i \lim_{p_1^2 \rightarrow 0}  \lim_{p_2^2 \rightarrow 0}  \lim_{p_3^2 \rightarrow 0} \, p_1^2 \, p_2^2 \, p_3^2 \, \frac{\delta}{\delta J(p_1)}  \frac{\delta}{\delta J(p_2)} \, \phi^{(1)}(-p_3) = -ig\, \del (p_1+ p_2 + p_3).
\end{equation}

It is useful to note that the operation of differentiating with respect to sources and multiplying by $p^2$ is equivalent to differentiating with respect to the solution $\phi^{(0)}$ of the zeroth order equation. Indeed, one can then consider the limit
\begin{equation}
\label{jonshell}
J(k) \rightarrow 0, \quad k^2 \rightarrow 0, \quad  j(k)\equiv - \frac{J(k)}{k^2} \neq 0.
\end{equation}
In this limit, the equation of motion becomes
\begin{equation}
\partial^2 \phi + \frac 12 g \phi^2 = 0,
\end{equation}
subject to the boundary condition that as $g \rightarrow 0$, the complete solution $\phi$ tends to a non-trivial solution $j(k)$ of the free equation (note that $j(k)$ has support on $k^2 = 0$). Working in this limit has the advantage that the background field is built from the objects $j(k)$ which have the interpretation of on-shell particles. The amplitude is given by
\begin{equation}
\mathcal{A}_n(p_1, \ldots, p_n) =-i \, \lim_{p_n^2 \rightarrow 0} \, p_n^2 \, \frac{\delta^{n-1} \phi^{(n-2)}(-p_n)}{\delta j(p_1)\delta j(p_2)\ldots\delta j(p_{n-1})}  .
\end{equation}
Note, however, that if we think of $\phi$ not as a background field, but simply as a generating functional for tree amplitudes, then we do not need to take the limit \eqref{jonshell}. The difference is that, in the latter case, we can take all legs to be off-shell, while if $\phi$ is a legitimate solution of the source-free field equation, then all but one legs (i.e. all legs representing sources) must be on-shell.

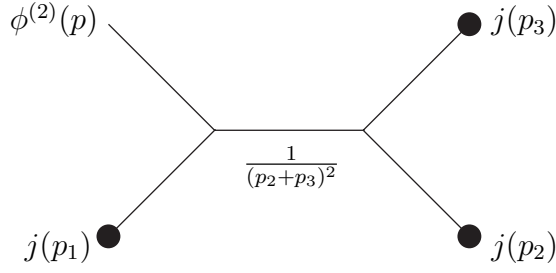
\begin{figure}
\centering
\includegraphics[scale=0.5]{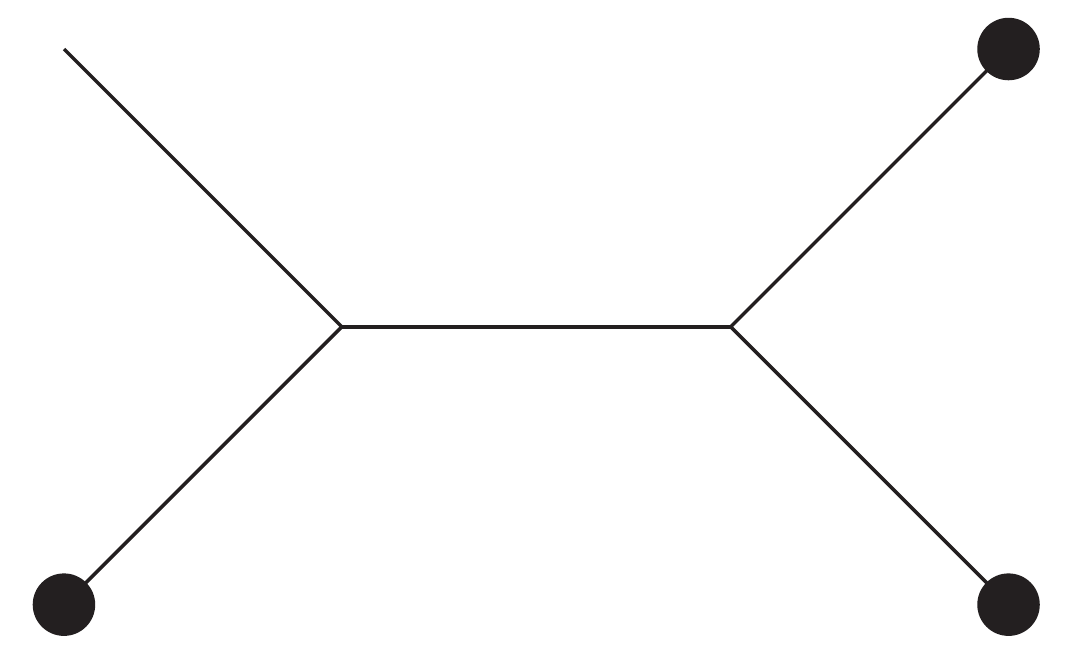}
\put(0,0){$j(p_2)$}
\put(-64,30){$\phi^{(2)}(p)$}
\put(0,30){$j(p_3)$}
\put(-62,0){$j(p_1)$}
\put(-33,11){$\frac{1}{(p_2+p_3)^2}$}
\caption{The second order correction $\phi^{(2)}(k)$ consists of an interaction between two particles, creating a disturbance which propagates before scattering against a third particle.}
\label{fig:scalar4}
\end{figure}

Let us continue to one more order in perturbation theory. The next correction $\phi^{(2)}(k)$ satisfies
\begin{equation}
k^2 \phi^{(2)} (k) = g \int \dd p_1 \dd p_2 \, \del (p_1 + p_2 - k) \, \phi^{(0)}(p_1) \phi^{(1)}(p_2).
\end{equation}
Inserting our expressions for $\phi^{(0)}$ and $\phi^{(1)}$, we find
\begin{equation}
\phi^{(2)}(k) = \frac 12 \int \dd p_1 \dd p_2 \dd p_3 \, \del (p_1 + p_2 + p_3 - k) \,  j(p_1) j(p_2) j(p_3)\, \frac{1}{k^2}\, g \, \frac{1}{(p_2 + p_3)^2} \, g
\end{equation}
The Feynman diagram corresponding to this expression is shown in Fig.~\ref{fig:scalar4}. The four-point amplitude is easily obtained by differentiation with respect to $j$ and amputating the single remaining off-shell leg.

\subsection{Off-shell spinor-like variables}

In the next sections, we will apply these techniques to Yang-Mills theory and to gravity. In anticipation of these applications, it is useful to describe some slight extension of the spinor-helicity method. Let us begin in familiar territory. We define Pauli matrices
\begin{singlespace}
\begin{equation}
\sigma^0 = \begin{pmatrix}
1 & 0 \\
0 & 1
\end{pmatrix}, \quad
\sigma^1= \begin{pmatrix}
0 & 1 \\
1 & 0
\end{pmatrix}, \quad
\sigma^2 = \begin{pmatrix}
0 & -i  \\
i & 0
\end{pmatrix}, \quad
\sigma^3 = \begin{pmatrix}
1 & 0 \\
0 & -1
\end{pmatrix}.
\vspace{0.5cm}
\end{equation}
\end{singlespace}
\noindent
We also define $\tilde \sigma^\mu = (\sigma^0, - \vec \sigma)$. These matrices satisfy the Clifford algebra
\begin{equation}
\sigma^\mu \tilde \sigma^\nu + \sigma^\nu \tilde \sigma^\mu = 2 \eta^{\mu \nu}.
\end{equation}
Given a momentum $p$, one can consider the object $p_{\alpha \dot \alpha} = p \cdot \sigma_{\alpha \dot \alpha}$ where $\alpha = 1, 2$ and $\dot \alpha = 1, 2$. If $p^2 = 0$ and $p\neq0$, the matrix $p_{\alpha \dot \alpha}$ has rank 1. Thus we may write $p_{\alpha \dot \alpha} = \lambda_\alpha \tilde \lambda_{\dot \alpha}$ where $\lambda$ and $\tilde \lambda$ are two component spinors. Notice that $\lambda$ and $\tilde \lambda$ are not uniquely defined; indeed, new spinors $\lambda' = \lambda / \zeta$ and $\tilde \lambda' = \zeta \tilde \lambda$ are just as good. We will exploit this freedom below.

The only Lorentz invariant associated with a single momentum $p_1$ is $p_1^2 = 0$. However, given two momenta $p_1$ and $p_2$, there are non-trivial invariants. We may form invariants from the spinors $\lambda_1$ and $\lambda_2$ associated with the two momenta. Lorentz transformations act on these spinors via a single copy of SU(2); therefore the object $\langle 12 \rangle \equiv \epsilon^{\alpha \beta} \lambda_{1 \alpha} \lambda_{2\beta}$ is a Lorentz invariant. Similarly, the object $[12] \equiv \epsilon^{\dot \alpha \dot \beta} \tilde \lambda_{1 \dot \alpha}\tilde \lambda_{2 \dot \beta}$ is an invariant. All other Lorentz invariants are built from these. In particular $(p_1 + p_2)^2 \equiv s_{12} = \langle 12 \rangle [12]$.

It will be convenient for us to define analogues of the spinorial products for off-shell momenta. In doing so, we follow the notation of \cite{Bardeen,Cangemi:1996rx}. We exploit the freedom to rescale the spinor $\lambda$ associated with an on-shell momentum $p$ to write $\lambda = (Q, 1)$. Note that $\lambda$ is now dimensionless; correspondingly, $\tilde \lambda$ has (mass) dimension 1. We then compute that $\tilde \lambda = (p_w, p_u)$ where
\begin{equation}
\label{coords}
u=t-z , \quad v=t+z , \quad w=x+i y.
\end{equation}
The spinor products are then given by
\begin{align}
[12] &\rightarrow X(p_1, p_2) \equiv p_{1 w} p_{2 u} - p_{1 u} p_{2 w}, \\
\langle 12 \rangle &\rightarrow Q(p_1, p_2) \equiv Q(p_1) - Q(p_2).
\end{align}
We will use the notation $X(p_1, p_2)$ and $Q(p_1, p_2)$ to emphasize that the spinor product is taken with this unusual rescaling of the spinors. The benefit is that the definition of $X$ extends to arbitrary off-shell momenta. We shall have more to say about the freedom to rescale spinors in the later sections of this paper.


\section{Self-dual Yang-Mills theory} 
\label{sec:sdym}

In this section, we begin our study of classical background fields. Since our goal is to investigate any possible BCJ-like structure of the fields, we choose to study the simplest non-trivial fields available. These are the self-dual solutions. Firstly, we consider the self-dual Yang-Mills (SDYM) equations in Minkowski spacetime,
\begin{equation}
\label{SDYM}
F_{\mu\nu} = \frac{i}{2} \varepsilon_{\mu\nu\rho\sigma} F^{\rho\sigma} .
\end{equation}
The gauge field is necessarily complexified, and the physical interpretation is that it is a configuration of positive helicity waves. Our set-up follows Bardeen and Cangemi \cite{Bardeen,Cangemi:1996rx}; see also \cite{Chalmers:1996rq}. We work with the coordinates \eqref{coords}, such that the metric is given by
\begin{eqnarray}
\label{mink}
ds^2 = du \, dv - dw \, d\bar{w} .
\end{eqnarray}
We choose the light-cone gauge, where $A_u=0$. The self-dual equations \eqref{SDYM} then imply\footnote{In our conventions, the field strengh is $F_{\mu\nu}=\partial_\mu A_\nu - \partial_\nu A_\mu - ig [A_\mu,A_\nu]$, and the structure constants of the Yang-Mills Lie algebra are defined by $[T^a, T^b] = i f^{abc}T^c$.}
\begin{equation}
\label{SDA}
A_w=0 , \quad A_v = -\frac{1}{4} \, \partial_w \Phi , \quad A_{\bar{w}} = -\frac{1}{4} \, \partial_u \Phi ,
\end{equation}
and also an equation of motion for the Lie algebra-valued scalar field $\Phi$,
\begin{equation}
\label{SDYMeom}
\partial^2 \Phi +i g [ \partial_w \Phi, \partial_u \Phi] =0,
\end{equation}
where $\partial^2 = 4\, (\partial_u \partial_v - \partial_w\partial_{\bar{w}})$ is the wave operator. Thus, our problem reduces to studying a scalar equation with a cubic coupling.

Following the discussion in the last section, let us now solve the scalar equation \eqref{SDYMeom} with the boundary condition that, when $g \rightarrow 0$, $\Phi(x) \rightarrow j(x)$. In momentum space, we can write this equation as
\begin{equation}
\label{SDYMeomk}
\Phi^a(k) = \frac{1}{2} \,g \int \dd p_1  \dd p_2  \, \frac{ F_{p_1 p_2}{}^k \, f^{b_1 b_2 a} }{k^2} \, \Phi^{b_1}(p_1)\Phi^{b_2}(p_2),
\end{equation}
where we have defined
\begin{equation}
\label{defF}
F_{p_1 p_2}{}^k \equiv \del(p_1+p_2-k) X(p_1,p_2) .
\end{equation}
We shall use an integral Einstein convention for the contraction of the indices of $F_{p_1 p_2}{}^k$,
\begin{align}
F_{p_1 q}{}^k \,F_{p_2 p_3}{}^q & \equiv \int \dd q \, \del(p_1+q-k) \, X(p_1,q) \, \del(p_2+p_3-q) \, X(p_2,p_3) \nonumber \\
& = \del(p_1+p_2+p_3-k) \, X(p_1,p_2+p_3) X(p_2,p_3) .
\end{align}
Moreover, we can lower and raise indices using
\begin{equation}
\delta^{p q} \equiv \del(p+q) = \delta_{p q}, \qquad \textrm{such that} \qquad  \delta_{p q} \delta^{q k} = \delta_p{}^k = \del(p-k).
\end{equation}
It is straightforward to see that $F^{p_1 p_2 p_3}=F_{p_1 p_2 p_3}$ is totally antisymmetric, e.g.
\begin{equation}
F^{p_1 p_2 p_3} = \del(p_1+p_2+p_3) X(p_1,-p_1-p_3) = - \del(p_1+p_2+p_3) X(p_1,p_3) = - F^{p_1 p_3 p_2} .
\end{equation}
Our notation is designed to emphasize the fact that the coefficients $F^{p_1 p_2 p_3}$ have the same algebraic properties as the structure constants $f^{abc}$. However, we will learn below that the significance is deeper, and that $F^{p_1 p_2 p_3}$ are, in fact, structure constants for a certain infinite-dimensional Lie algebra.

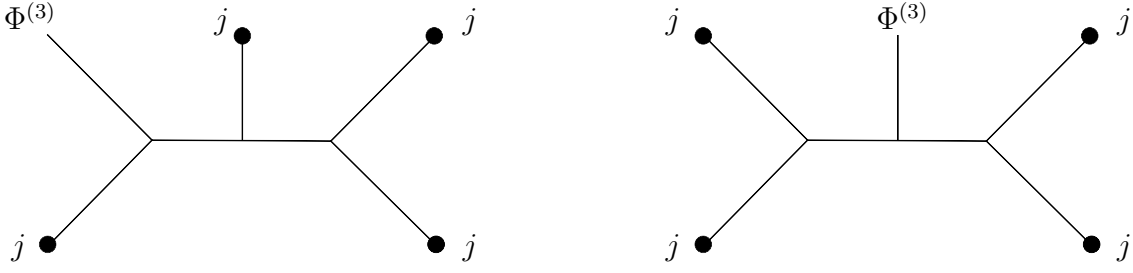
\begin{figure}
\centering
\includegraphics[scale=0.6]{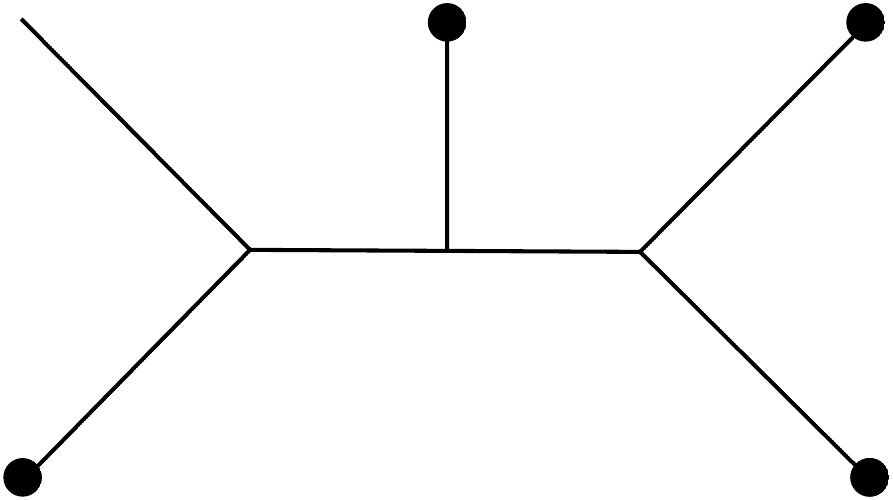} \hspace{3cm} \includegraphics[scale=0.6]{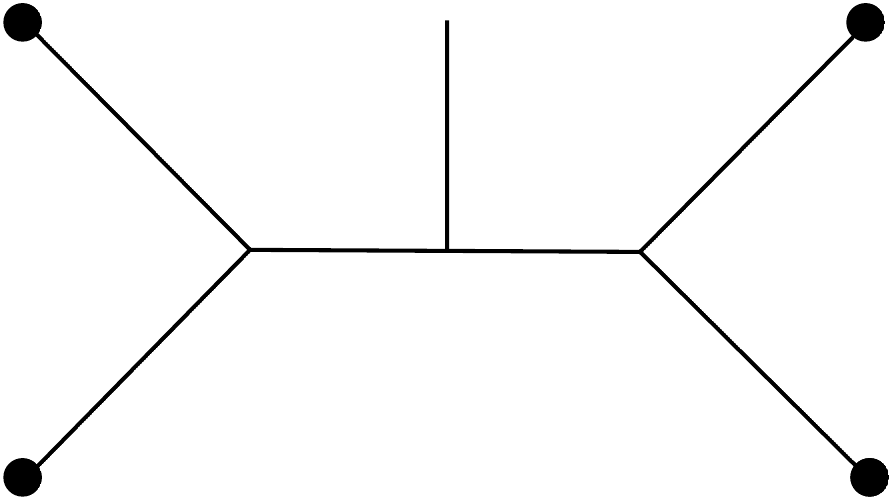}
\put(2,0){$j$} \put(-58,0){$j$}
\put(2,30){$j$} \put(-58,30){$j$}
\put(-30,30){$\Phi^{(3)}$}
\put(-146,30){$\Phi^{(3)}$}
\put(-85,0){$j$} \put(-145,0){$j$}
\put(-85,30){$j$} \put(-118,30){$j$}
\caption{The third order correction $\Phi^{(3)}(k)$, in \eqref{Phi3}, consists of two different terms. The first one corresponds to the Feynman diagram on the left, and the second one to the diagram on the right.}
\label{fig:sdym5}
\end{figure}

Equipped with this formalism, we solve the equation of motion for $\Phi^a$ \eqref{SDYMeomk} iteratively, as a series expansion in the coupling constant $g$. The first few terms are
\begin{align}
\Phi^{(0)a} (k) &= j^a(k) , \\
\Phi^{(1)a} (k) &= \frac{1}{2} \,g \int \dd p_1  \dd p_2 \, \frac{ F_{p_1 p_2}{}^k \, f^{b_1 b_2 a}}{k^2} \,j^{b_1}(p_1) j^{b_2}(p_2), \\
\Phi^{(2)a} (k) &= \frac{1}{2} \,g^2 \int \dd p_1 \dd p_2 \dd p_3  \,
\frac{F_{p_1 q}{}^k \,F_{p_2 p_3}{}^q \, f^{b_1 c a} f^{b_2 b_3 c}}{k^2(p_2+p_3)^2} \, j^{b_1}(p_1) j^{b_2}(p_2) j^{b_3}(p_3) .
\label{Phi2}
\end{align}
As one would expect, different terms appear at higher orders,
\begin{align}
\Phi^{(3)a} (k) &=\frac{1}{2} \,g^3 \int \dd p_1 \dd p_2 \dd p_3  \, 
\Bigg( \frac{F_{p_1 q_1}{}^k \, F_{p_2 q_2}{}^{q_1} \, F_{p_3 p_4}{}^{q_2} \, f^{b_1 c_1 a} f^{b_2 c_2 c_1} f^{b_3 b_4 c_2}}{k^2(p_2+p_3+p_4)^2(p_3+p_4)^2} + \phantom{aaa} \nonumber \\
& \phantom{aaaa} + \frac{F_{q_1 q_2}{}^k \, F_{p_1 p_2}{}^{q_1} F_{p_3 p_4}{}^{q_2} \,f^{c_1 c_2 a} f^{b_1 b_2 c_1} f^{b_3 b_4 c_2}}{4\,k^2(p_1+p_2)^2(p_3+p_4)^2} \Bigg) \, j^{b_1}(p_1) j^{b_2}(p_2) j^{b_3}(p_3)j^{b_4}(p_4).
\label{Phi3}
\end{align}
The two terms correspond to two types of Feynman diagrams, represented in Fig.~\ref{fig:sdym5}.

The way in which the kinematic factors $F^{p_1 p_2 p_3}$ mirror the commutators is a consequence of equation \eqref{SDYMeomk}. Indeed, the colour/kinematics correspondence is simply
\begin{align}
F^{p_1 p_2 p_3}  \quad & \longleftrightarrow \quad f^{abc}.
\label{kincol}
\end{align}
In the same way that each vertex is dressed by a $f^{abc}$ factor, it is also dressed by a kinematic $F^{p_1 p_2 p_3}$ factor. Finally, the kinematic Jacobi identity is given by
\begin{align}
F_{p_1 p_2}{}^q \, F_{p_3 q}{}^k + F_{p_2 p_3}{}^q \, F_{p_1 q}{}^k + F_{p_3 p_1}{}^q \, F_{p_2 q}{}^k =0,
\label{kinjacobi}
\end{align}
which is a consequence of\footnote{On-shell, this identity can be understood as a special case of the Schouten identity.}
\begin{align}
X(k_1,k_2) X(k_3,k_1+k_2) + X(k_2,k_3) X(k_1,k_2+k_3) + X(k_3,k_1) X(k_2,k_3+k_1) = 0 .
\label{kinjacobiX}
\end{align}

The relation that we found here is exactly of the type that BCJ pointed out for Yang-Mills scattering amplitudes \cite{Bern:2008qj}, but now in the context of classical background solutions. The obvious question now is: what is the algebra whose structure constants are the $F^{p_1 p_2 p_3}$ factors? The answer appears naturally once we look at the gravity case.

For completeness, let us remark that there is a simple expression valid to all orders \cite{Cangemi:1996rx},
\begin{align}
\Phi^{(n)} (k) &= (ig)^n \int \dd p_1  \dd p_2 \ldots \dd p_{n+1} \, \del(p_1+p_2+\ldots +p_{n+1}-k) \nonumber \\ & \phantom{aaaaa}  \times  j(p_1)j(p_2)\ldots j(p_{n+1}) \, Q(p_1,p_2)^{-1}Q(p_2,p_3)^{-1} \ldots Q(p_n,p_{n+1})^{-1} .
\label{Phinall}
\end{align}
However, this form is not convenient for our purposes, since the Feynman diagram expansion is not explicit. Moreover, unlike the preceding discussion, this expression is only valid if $j(k)$ has support on $k^2=0$, i.e. if the legs representing sources correspond to on-shell particles.


\section{Self-dual gravity}
\label{sec:sdgr}

We consider now self-dual gravity (SDG) with Lorentzian signature, using an approach analogous to the one we used for SDYM. Previously, Mason and Skinner~\cite{Mason:2008jy} have computed the MHV amplitudes in gravity by perturbing the classical, self-dual background field, in work analogous to that of Bardeen and Cangemi \cite{Bardeen,Cangemi:1996rx} on self-dual Yang-Mills theory. Our focus here will be on the background field itself rather than perturbations around it. The self-dual equations are
\begin{equation}
\label{SDG}
R_{\mu\nu\lambda\delta} = \frac{i}{2} \varepsilon_{\mu\nu\rho\sigma} R^{\rho\sigma}_{\phantom{\rho\sigma}\lambda\delta} .
\end{equation}
Inspired by the gauge field \eqref{SDA}, we try as a solution the metric
\begin{equation}
g_{\mu\nu}=\eta_{\mu\nu}+\kappa \,h_{\mu\nu} ,
\end{equation}
where $\kappa$ is the gravitational coupling, $\eta_{\mu\nu}$ is the Minkowski metric \eqref{mink}, and the non-vanishing components of $h_{\mu\nu}$ are
\begin{equation}
\label{SDh}
h_{vv} = -\frac{1}{4} \, \partial_w^2 \phi , \quad h_{\bar{w}\bar{w}} = -\frac{1}{4} \, \partial_u^2 \phi , \quad h_{v\bar{w}} = h_{\bar{w}v} = -\frac{1}{4} \, \partial_w \partial_u \phi .
\end{equation}
The SDG equations \eqref{SDG} then imply that the scalar field $\phi$ obeys\footnote{The possible contributions to the right-hand-side of \eqref{SDGeom0} can be absorbed by a redefinition of $\phi$.}
\begin{equation}
\label{SDGeom0}
\partial^2 \phi + \kappa \big( (\partial_w^2 \phi) \, (\partial_u^2 \phi) - (\partial_w \partial_u \phi )^2 \big)=  0,
\end{equation}
where $\partial^2$ denotes the Minkowski space wave operator. It turns out that this equation was first obtained by Pleba\~nski \cite{Plebanski:1975wn}, and that it allows for the most general solution of SDG.

The resemblance between SDYM and SDG becomes even more striking if we rewrite \eqref{SDGeom0} as
\begin{equation}
\label{SDGeom}
\partial^2 \phi +\kappa \{ \partial_w \phi, \partial_u \phi \} = 0,
\end{equation}
which should be compared to \eqref{SDYMeom}. We introduced here the Poisson bracket
\begin{equation}
\label{bracket}
\{ f, g \} \equiv (\partial_w f)\,( \partial_u g) - (\partial_u f)\,( \partial_w g) ,
\end{equation}
from which we construct the Poisson algebra
\begin{equation}
\label{Palgebra}
\{ e^{-ik_1 \cdot x} , e^{-ik_2 \cdot x} \} = -X(k_1,k_2) \,e^{-i(k_1+k_2) \cdot x}.
\end{equation}
This is the kinematic algebra that we were looking for in the last section. It is the Poisson version of the algebra of area-preserving diffeomorphisms of $w$ and $u$. To see this, consider a diffeomorphism $w \to w'(w,u)$, $u \to u'(w,u)$. This transformation preserves the Poisson bracket \eqref{bracket} if and only if it has a unit Jacobian, i.e. it is area-preserving. The infinitesimal generators of the diffeomorphisms are
\begin{equation}
L_k = e^{-ik \cdot x} (-k_w \partial_u + k_u \partial_w ) .
\end{equation}
The Lie algebra is
\begin{equation}
[ L_{p_1} , L_{p_2} ] = i X(p_1,p_2) L_{p_1+p_2} = i F_{p_1 p_2}{}^k L_k,
\end{equation}
and the Jacobi identity was given in \eqref{kinjacobi}.

We have seen that this algebra is the kinematic analogue of the Yang-Mills Lie algebra. It is interesting to note that there is a correspondence between the Lie algebra of area-preserving diffeomorphisms of $S^2$ and the Lie algebra of the generators of SU($N$) in the planar limit $N\to \infty$, in the sense that there exists an appropriate basis such that the structure constants are the same \cite{Hoppe}.

Let us now proceed, as in the SDYM case, to solve this gravitational equation \eqref{SDGeom}. In momentum space, the equation can be written as
\begin{equation}
\label{SDGeomk}
\phi(k) =  \frac{1}{2} \,\kappa \int \dd p_1  \dd p_2 \, \frac{X(p_1, p_2) F_{p_1 p_2}^{\phantom{p_1 p_2}k}}{k^2} \, \phi(p_1) \phi(p_2),
\end{equation}
We again take $\phi^{(0)}(k) = j(k)$ to have support on the light cone. To the first few orders in $\kappa$, the solution is
\begin{align}
\phi^{(0)} (k) &= j(k) , \\
\phi^{(1)} (k) &= \frac{1}{2} \,\kappa \int \dd p_1  \dd p_2 \, \frac{X(p_1, p_2) F_{p_1 p_2}^{\phantom{p_1 p_2}k}}{k^2} \,j(p_1) j(p_2), \\
\phi^{(2)} (k) &= \frac{1}{2} \,\kappa^2 \int \dd p_1  \dd p_2  \dd p_3 \,
\frac{X(p_1, q) F_{p_1 q}^{\phantom{p_1 q}k} X(p_2, p_3) F_{p_2 p_3}^{\phantom{p_2 p_3}q}}{k^2(p_2+p_3)^2}\,  j(p_1) j(p_2) j(p_3).
\label{phi2}
\end{align}

Let us explore the relationship between the expressions (\ref{SDYMeomk}-\ref{Phi2}) and (\ref{SDGeomk}-\ref{phi2}). Indeed, it is clear that one can deduce the gravitational expressions from the Yang-Mills cases by replacing the SU(N) structure constants $f^{abc}$ by appropriate factors of $X$. These factors of $X$ are, in turn, related to the structure constants $F$ of the kinematic algebra. However, the relationship is not given by $f \rightarrow F$ because this would involve squaring a delta function. One algorithm for deducing the gravitational expressions from the Yang-Mills formulae involves extracting the overall momentum conserving delta function, and then following the BCJ procedure of identifying a kinematic numerator which is to be squared. Let us illustrate this at the level of the second corrections, $\Phi^{(2)a}$ and $\phi^{(2)}$. Beginning with the Yang-Mills formula, we have
\begin{align}
\Phi^{(2)a}(k)=& \frac{1}{2} \,g^2 \int \dd p_1 \dd p_2 \dd p_3 \, \del (p_1+p_2+p_3-k) \nonumber \\
& \phantom{aaaaaaaaa} \times \frac{X(p_1, p_2 + p_3)  X(p_2, p_3) \, f^{b_1 c a} f^{b_2 b_3 c}}{k^2(p_2+p_3)^2} \, j^{b_1}(p_1) j^{b_2}(p_2) j^{b_3}(p_3).
\end{align}
The corresponding off-shell tree-level correlator -- we will refer to this object as an off-shell ``amplitude'' -- is obtained by differentiating with respect to the $j^a(p)$ and amputating the final leg. The expression now becomes
\begin{equation}
\mathcal{A}^{b_1 b_2 b_3 a}(p_1, p_2, p_3,-k) = \frac{1}{2} \,g^2 \,
\frac{X(p_1, p_2+p_3)  X(p_2, p_3) \, f^{b_1 c a} f^{b_2 b_3 c}}{(p_2+p_3)^2} \, \del (p_1+p_2+p_3-k) + \cdots,
\end{equation}
where the dots indicate two other channels. The BCJ procedure now identifies the kinematic numerator as 
\begin{equation}
n_t = X(p_1, p_2 + p_3)  X(p_2, p_3) 
\end{equation}
This is the object which must replace the colour factor $f^{b_1 c a} f^{b_2 b_3 c}$ to deduce the gravitational perturbation, as one easily checks. Moreover, as expression \eqref{kinjacobiX} shows, these objects satisfy a Jacobi identity. Of course, we could have deduced this numerator directly from the background field perturbation; we have introduced the off-shell ``amplitude'' merely to illustrate that the procedure is equivalent to the BCJ procedure.


\section{Yang-Mills theory}
\label{sec:full}

We have seen that there is a remarkably close connection between the equations governing self-dual background fields in Yang-Mills theory and in gravity. We will now consider the full Yang-Mills theory in the light of the kinematic algebra present in the self-dual sector.

There is one non-trivial scattering amplitude in the self-dual theory, which is the three-point amplitude (evaluated for complex momenta). In fact, this three-point amplitude is the same as the amplitude in the full theory. Let us briefly explain why this is so. In Yang-Mills theory, the three-point $++-$ scattering amplitude is given by
\begin{equation}
\mathcal{A}_3^{abc}(p_1,p_2,p_3) =- i g \,\frac{[12]^3}{[23][31]}\, \tilde{f}^{abc} \,\del(p_1+p_2+p_3),
\end{equation}
where $\tilde{f}^{abc}\equiv\sqrt{2}f^{abc}$. It is convenient to rescale the spinors as described in section~\ref{sec:bfReview} to make contact with the invariants $X(p_i,p_j)$. In terms of a scaling factor $\zeta_i$ for $i=1,2,3$, we have
\begin{equation}
[12] = \zeta_1 \zeta_2 \,X(p_1,p_2), \quad
[23] = \zeta_2 \zeta_3 \,X(p_2,p_3), \quad
[31] = \zeta_3 \zeta_1 \,X(p_3,p_1),
\end{equation}
so that the three-point amplitude can be written as
\begin{align}
\mathcal{A}_3^{abc}(p_1,p_2,p_3) &= -ig \,\frac{\zeta_1^2 \zeta_2^2}{\zeta_3^2} \, \frac{X(p_1,p_2)^3}{X(p_2,p_3)X(p_3,p_1)} \tilde{f}^{abc} \, \del(p_1+p_2+p_3) \\
&= -ig \, \frac{\zeta_1^2 \zeta_2^2}{\zeta_3^2} \, X(p_1,p_2) \tilde{f}^{abc} \, \del(p_1+p_2+p_3) \\
&\to -ig \, F_{p_1 p_2 p_3} \tilde{f}^{abc} ,
\label{A3Ff}
\end{align}
where in the last line we redefined our basis of polarization vectors to absorb the factors $\zeta_i$. Then this Yang-Mills amplitude is given by the same expression as in the self-dual sector. It is important to note that, while we started from a spinor-helicity on-shell expression, the final result \eqref{A3Ff} can be obtained off-shell.

Similar comments hold for the $--+$ Yang-Mills amplitude; a basis of polarization vectors can be found so that the YM amplitude is given by the three-point amplitude in the anti-self-dual sector. It is convenient to define
\begin{equation}
\Fb_{p_1 p_2}{}^k \equiv \del(p_1+p_2-k) \,\Xb(p_1,p_2), \qquad \textrm{where} \quad
\Xb(p_1,p_2) \equiv p_{1\wb}p_{2u}-p_{1u}p_{2\wb}.
\end{equation}
Then the anti-self-dual three-point amplitude (and the $--+$ YM amplitude) is simply
\begin{equation}
\mathcal{A}_3^{abc}(p_1,p_2,p_3) = ig\, \bar X(p_1,p_2) \tilde{f}^{abc} \,\del(p_1+p_2+p_3) = ig\, \bar F_{p_1 p_2 p_3} \tilde{f}^{abc},
\end{equation}
where $\bar F_{p_1 p_2 p_3} = \bar X(p_1,p_2)  \del(p_1+p_2+p_3)$ are the structure constants for the anti-self-dual kinematic algebra. Note that, if $p_1$ and $p_2$ are on-shell, then the quantities $X(p_1,p_2)$ and $\Xb(p_1,p_2)$ are related by
\begin{equation}
X(p_1,p_2)\,\Xb(p_1,p_2) = -\frac{1}{4}\, p_{1u}p_{2u} \,s_{12}.
\end{equation}

In the gravitational case, it is also true that a polarization basis can be chosen so that the three-point functions in the full theory are identical to the three-point functions we have discussed in the self-dual sector.

\subsection{Chalmers-Siegel Theory and the Four-Point Amplitude}

We have seen that the kinematic algebra is present in the full theory, at least at the level of the three-point amplitudes. However, it seems that two algebras appear: one for the MHV three-point amplitude, and another for the $\overline{\mathrm{MHV}}$ case. To study how these algebras interact, it is convenient to turn to the Chalmers-Siegel formulation of Yang-Mills theory in light-cone gauge~\cite{Chalmers:1998jb}. Their action describes Yang-Mills theory as a theory of interacting positive and negative helicity degrees of freedom, and therefore is very helpful to connect the full theory with the discussion of Section~\ref{sec:sdym}. We will see here how the colour/kinematics duality for the self-dual (or anti-self-dual) sector underlies the BCJ relations in the MHV sector of Yang-Mills theory.

As before, we take the light-cone gauge
\begin{equation}
A_\mu=(0,A_v,\Ab,A),
\end{equation}
where we again work with the coordinates \eqref{coords}, and denote $\Ab\equiv A_w$, $A \equiv A_{\bar{w}}$, to avoid the proliferation of indices. Positive helicity particles correspond to $\Ab=0$, and negative helicity particles to $A=0$. The great simplification of the light-cone gauge is that now $A_v$ appears quadratically in the Yang-Mills Lagrangian and can be integrated out. We are left with
\begin{equation}
\label{CSlag}
{\mathcal L}= \textrm{tr}\Big\{\;  \frac{1}{4} \Ab \,\partial^2 A -ig \Big(\frac{\partial_w}{\partial_u}A\Big) [A,\partial_u \Ab] 
- ig \Big(\frac{\partial_{\wb}}{\partial_u}\Ab\Big) [\Ab,\partial_u A] -g^2 [A,\partial_u \Ab]\frac{1}{\partial_u^2} [\Ab,\partial_u A] \;\Big\}.
\end{equation}
The equations of motion are
\begin{align}
\frac{1}{4}\, \partial^2 A =& -i g\,[\partial_w A,A] + i g\,\frac{\partial_{\wb}}{\partial_u} [\Ab,\partial_u A]
-i g\,[ \frac{\partial_w}{\partial_u}A + \frac{\partial_{\wb}}{\partial_u}\Ab ,\partial_u A] \nonumber \\
& + g^2 [A,\frac{1}{\partial_u}[\Ab,\partial_u A]] +g^2 [\partial_u A,\frac{1}{\partial_u^2} ([A,\partial_u \Ab]+[\Ab,\partial_u A]) ]
\end{align}
and the conjugate equation for $\Ab$. One can easily check that equation \eqref{SDYMeom} is recovered for the choice $\Ab=0$, $A=-\partial_u \Phi/4$. In momentum space, we have
\begin{align}
&A^a(k) = -i g\, \int \dd p_1  \dd p_2 \, \Big\{ 2 \, \frac{ k_u }{p_{1u}p_{2u}} \, \frac{ F_{p_1 p_2}{}^k \, f^{b_1 b_2 a} }{k^2} \, A^{b_1}(p_1) A^{b_2}(p_2) \nonumber \\ & \hspace{3cm} 
+  4 \, \frac{ p_{1u} }{p_{2u}k_u} \, \frac{ \Fb_{p_1 p_2}{}^k \, f^{b_1 b_2 a} }{k^2} \, A^{b_1}(p_1) \Ab^{b_2}(p_2) \Big\}  \nonumber \\
& + 4 g^2\,\int \dd p_1  \dd p_2 \dd p_3\, \del(p_1+p_2+p_3-k)\, \frac{-p_{2u}k_u+p_{1u}p_{3u}}{(p_{2u}+p_{3u})^2}\,
\frac{  f^{b_1 c a} f^{b_2 b_3 c} }{k^2} \,  A^{b_1}(p_1) A^{b_2}(p_2) \Ab^{b_3}(p_3) .
\label{fullYMeomk}
\end{align}
There are now two more vertices with respect to the self-dual case: the cubic vertex from the anti-self-dual theory and the four-point vertex.

We proceed as in the past sections to construct the solution perturbatively, starting from
\begin{equation}
A^{(0)a} (k) = \eta^a (k), \qquad \Ab^{(0)a} (k) = \etab^a (k).
\end{equation}
We can then obtain the tree-level $++--$ (off-shell) amplitude
\begin{equation}
{\mathcal A}(1^+ 2^+ 3^- 4^-) = 
\varepsilon^+_w(p_1)\varepsilon^+_w(p_2)\varepsilon^-_{\wb}(p_3)\varepsilon^-_{\wb}(p_4)
\,p_4^2\,  \frac{\delta}{\delta \eta^{b_1}(p_1)} \frac{\delta}{\delta \eta^{b_2}(p_2)} \frac{\delta}{\delta \etab^{b_3}(p_3)} A^{(2)b_4}(-p_4).
\end{equation}
The on-shell amplitude is computed by taking the limit $p_i^2 \to 0$, for $i=1,2,3,4$.
There are several Feynman diagrams contributing to this amplitude, including the four-point vertex diagrams. We choose a gauge by taking $p_4$ to be a reference leg in the sense of Chalmers and Siegel \cite{Chalmers:1998jb}. Thus we take the limit $p_{4u}\to 0$,\footnote{We consider complex momenta, and thus this condition does not imply that $p_{4}$ is on-shell.} so that the four-point vertex is eliminated.
\begin{align}
{\mathcal A}&(1^+ 2^+ 3^- 4^-) = g^2 \,
\varepsilon^+_w(p_1)\varepsilon^+_w(p_2)\varepsilon^-_{\wb}(p_3)\varepsilon^-_{\wb}(p_4) \, \frac{16}{p_{4u}} \,
\frac{p_{3u}}{p_{1u}p_{2u}}\, \times \nonumber \\
& \Bigg\{ \frac{ F_{p_1 p_2 q} \, \Fb_{p_3 p_4}{}^q \, f^{b_1 b_2 c}  f^{b_3 b_4 c} }{{s}}  + \frac{ F_{p_2 p_3}{}^q \, \Fb_{p_1 p_4 q} \, f^{b_2 b_3 c} f^{b_1 b_4 c} }{{u}}
+ \frac{ F_{p_3 p_1}{}^q \, \Fb_{p_2 p_4 q} \,  f^{b_3 b_1 c} f^{b_2 b_4 c} }{{t}} 
\Bigg\},
\label{current}
\end{align}
where we introduced the Mandelstam variables,
\begin{equation}
{s} = (p_1+p_2)^2, \qquad {t} = (p_1+p_3)^2, \qquad {u} = (p_1+p_4)^2.
\end{equation}
There is no divergence in the amplitude as $p_{4u}\to 0$ since it is cancelled by $\varepsilon^-_{\wb}(p_4) \propto p_{4u}$.

The form of the amplitude \eqref{current} indicates that the BCJ kinematic numerators are
\begin{equation}
n_{s} =\alpha\,  X(p_1,p_2) \, \Xb(p_3,p_4), \quad 
n_{t} = \alpha\,  X(p_2,p_3) \, \Xb(p_1,p_4), \quad 
n_{u} = \alpha\, X(p_3,p_1) \, \Xb(p_2,p_4),
\end{equation}
where $\alpha$ is a proportionality factor. Because we now have both $X$ and $\Xb$, it may not seem that
\begin{equation}
\label{BCJJacobi}
n_{s} + n_{t} + n_{u} = 0,
\end{equation}
but recall that we set $p_{4u}=0$, so that
\begin{equation}
n_{s} =\alpha' \,  X(p_1,p_2) \, X(p_3,p_4), \quad 
n_{t} = \alpha'\,  X(p_2,p_3) \, X(p_1,p_4), \quad 
n_{u} = \alpha'\, X(p_3,p_1) \, X(p_2,p_4),
\end{equation}
where $\alpha'= \alpha \,p_{4\wb}/p_{4w} $. Then the relation \eqref{BCJJacobi} is nothing but the Jacobi identity \eqref{kinjacobiX} of the self-dual theory. We stress that all external legs can be taken to be off-shell.

In order to go on-shell, we can take $p_{4u}=p_{4w}=0$ and write
\begin{equation}
n_{s} =-\alpha\,p_{4\wb}\, X(p_1,p_2) \, p_{3u}, \quad 
n_{t} =- \alpha\,p_{4\wb}\,  X(p_2,p_3) \, p_{1u}, \quad 
n_{u} = -\alpha\,p_{4\wb} \,X(p_3,p_1) \, p_{2u},
\end{equation}
so that \eqref{BCJJacobi} now becomes
\begin{equation}
X(p_1,p_2) \, p_{3u}+ X(p_2,p_3) \, p_{1u} + X(p_3,p_1) \, p_{2u} = 0.
\end{equation}
Ref.~\cite{Chalmers:1998jb} details the procedure by which the amplitude can be put in the standard spinor-helicity form, so we will not address this.

\subsection{Higher-point MHV amplitudes \label{subsec:higherpoint}}

In our study of the four-point Yang-Mills amplitude, we found it helpful to make the reference leg choice $p_{4u} \rightarrow 0$. Any reference leg must be attached to a three-point vertex of the same self-duality; that is, a negative helicity reference leg must be attached to a $+--$ vertex. In the case of the four-point function, this choice of reference leads to the off-shell amplitude being written in terms of products of a single $+--$ vertex and a single $++-$ vertex. Moreover, with this choice both vertices are proportional to the structure constants of the self-dual kinematic algebra. We shall again exploit this simplification to clarify the relationship between the Jacobi identities satisfied by the MHV amplitudes and our kinematic algebra.

We will show that all the Feynman diagrams contributing to an MHV $n$-point amplitude, containing $n-2$ particles of positive helicity and 2 particles of negative helicity, contain only three-point vertices. By choosing a negative helicity line to be the reference leg, we ensure that the vertex attached to that leg is a $--+$ vertex, while all the other vertices are $++-$ vertices. To see this, let $n_+$ be the number of $++-$ three-point vertices in a given Feynman diagram, while $n_-$ is the number of $--+$ vertices. Similarly, let $n_4$ be the number of four-point vertices. We suppose there are $I$ internal lines at $n$ points. It is straightforward to establish a number of identities relating these quantities. Counting the powers of the coupling $g$, we find that
\begin{equation}
\label{ngym}
n - 2 = n_+ + n_- + 2 n_4.
\end{equation}
Meanwhile, the positive helicity ends of internal or external lines must terminate in the vertices, and similarly for the negative helicity lines:
\begin{align}
n-2 + I &= 2 n_+ + n_- + 2 n_4, \\
2 + I &= n_+ + 2 n_- + 2 n_4.
\label{nmym}
\end{align}
Manipulation of these identities leads to the conclusion that there is a strong constraint on the sum of the number of $--+$ vertices and four-point vertices:
\begin{equation}
n_- + n_4 = 1.
\label{nfinalym}
\end{equation}
Since we choose one negative helicity line as a reference, it follows that $n_- = 1$ and so $n_4 = 0$. Therefore, the MHV amplitude is computed from cubic diagrams.
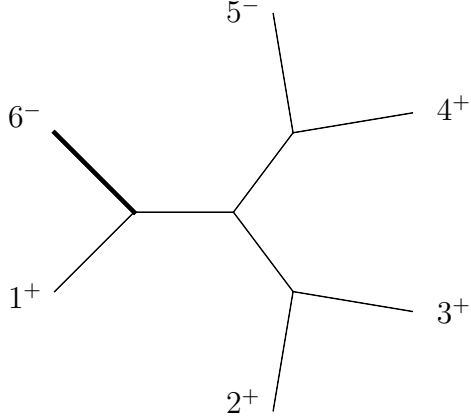
\begin{figure}
\centering
\includegraphics[scale=0.8]{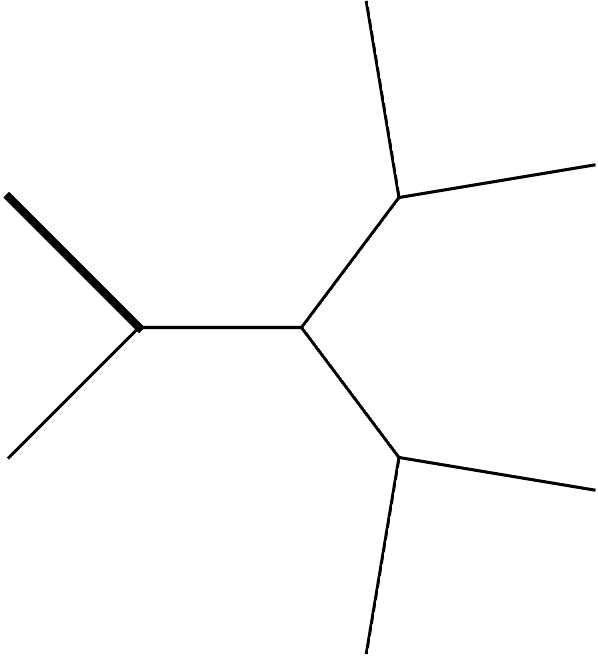}
\put(3,12){$3^+$}
\put(-25,0){$2^+$}
\put(-25,52){$5^-$}
\put(-54,38){$6^-$}
\put(3,39){$4^+$}
\put(-54,14){$1^+$}
\caption{Example of a diagram contributing to the 6-point MHV amplitude. The thick line represents the reference leg; the vertex attached to it is of the type $--+$, while the other three vertices are of the type $++-$.}
\label{fig:6pt}
\end{figure}
As an example, we present in Fig.~\ref{fig:6pt} a diagram contributing to the 6-point MHV amplitude. The negative helicity line represented in bold is the reference leg, which is connected to the only $--+$ vertex. The contribution from this diagram is proportional to
\begin{equation}
\frac{ \Fb_{p_1 p_6}{}^{q_1} \, F_{p_2 p_3}{}^{q_2} \, F_{p_4 p_5}{}^{q_3} \, F_{q_1 q_2 q_3} \,
 f^{b_1 b_6 c_1}  f^{b_2 b_3 c_2} f^{b_4 b_5 c_3}  f^{c_1 c_2 c_3}}{(p_1+p_6)^2(p_2+p_3)^2(p_4+p_5)^2}.
\end{equation}

Just as in the four-point case, the choice of reference line attached to the single $--+$ vertex in the amplitude means that the quantity $\bar X$ appearing in the vertex is proportional to $X$. Expressed in this way, the Jacobi relations satisfied by the numerators are nothing but the Jacobi relations of the self-dual kinematic algebra.


\section{Gravity}
\label{sec:gravity}

In this section, we will see how the kinematic algebra arises for gravity, and confirm that it indeed corresponds to the BCJ ``square'' of the Yang-Mills case. We need a formulation of the theory in terms of positive and negative helicity particles, analogous to the Chalmers-Siegel action, in order to make a connection with Section~\ref{sec:sdgr}. Fortunately, such a Lagrangian was obtained by Ananth {\it et al} \cite{Ananth:2006fh}; see also \cite{Ananth:2007zy}.

The Lagrangian is obtained from the Einstein-Hilbert Lagrangian in the following way (see Appendix~C of \cite{Ananth:2006fh} for a detailed derivation). Take the light-cone gauge for the metric,
\begin{equation}
g_{uu}=g_{ui}=0,
\end{equation}
where $i=1,2$ denotes the $x$ and $y$ Cartesian coordinates. Parameterize the remaining components as
\begin{equation}
g_{uv}=\frac{1}{2} \,e^{\psi/2}, \qquad g_{ij}=-e^{\psi/2} \gamma_{ij},
\end{equation}
where $\gamma_{ij}$ is a symmetric $2 \times 2$ matrix of unit determinant. Then the Einstein constraints $R_{uu}=R_{ui}=0$ allow us to write the Lagrangian in terms of $\gamma_{ij}$ only. The perturbative expansion in terms of positive helicity gravitons $h$ and negative helicity gravitons $\hb$ is performed by taking
\begin{equation}
\gamma_{ij}=\left(\mathrm{e}^{\kappa H}\right)_{ij}\ ,\qquad
H=\frac{1}{\sqrt 2} \begin{pmatrix} h+\hb & i(h-\hb) \\ i(h-\hb) &-(h+\hb) \end{pmatrix} .
\end{equation}
Expanding in the coupling $\kappa$, we obtain the Lagrangian\footnote{As described in \cite{Ananth:2006fh}, a field redefinition which removes occurences of $\partial_v$ from the four-point interaction has been performed; this is irrelevant for our purposes.}
\begin{align}
& {\cal L} = \frac{1}{4} \, \hb \,\partial^2 h
+ \kappa \; \hb\, \partial_u^2\left( -h\,\frac{\partial_w^2}{\partial_u^2} h +\left(\frac{\partial_w}{\partial_u} h\right)^2  \right)
+ \kappa \; h\, \partial_u^2\left( -\hb\,\frac{\partial_{\wb}^2}{\partial_u^2} \hb + \left(\frac{\partial_{\wb}}{\partial_u} \hb \right)^2 \right) \nonumber \\
& + \kappa^2 \Bigg[
\frac{1}{\partial_u^2} \big(\partial_u h \,\partial_u \hb \big)\, \frac{\pwwb}{\partial_u^2} \big(\partial_u h\, \partial_u \hb \big)
+\frac{1}{\partial_u^3} \big(\partial_u h \,\partial_u \hb \big)\, \left(\pwwb h\,\partial_u\hb+\partial_uh \,\pwwb \hb\right)  \nonumber \\
& -\frac{1}{\partial_u^2} \big(\partial_u h\, \partial_u \hb \big) \,\left(2\,\pwwb h\,\hb+2\,h \,\pwwb \hb
+9\,\partial_w h \, \partial_{\wb}\hb+\partial_{\wb} h\, \partial_w\hb -\frac{\pwwb}{\partial_u}h\,\partial_u\hb-\partial_u h \,\frac{\pwwb}{\partial_u} \hb \right) \nonumber  \\
&-2\frac{1}{\partial_u}\big(2\,\partial_w h\,\partial_u\hb+h \,\pwu\hb-\pwu h\,\hb\big)\,h\,\partial_{\wb}\hb
-2\frac{1}{\partial_u}\big(2\,\partial_u h\,\partial_{\wb}\hb+\pwbu h\,\hb-h\,\pwbu\hb \big)\,\partial_w h\,\hb \nonumber  \\
& -\frac{1}{\partial_u} \big(2\,\partial_w h\,\partial_u\hb+h\,\pwu\hb-\pwu h\,\hb \big)
\frac{1}{\partial_u}\big(2\,\partial_u h\,\partial_{\wb}\hb + \pwbu h\,\hb-h\,\pwbu \hb \big) \nonumber \\
& -h\,\hb\,\bigg(\pwwb h\,\hb+h\,\pwwb\hb+2\,\partial_w h\, \partial_{\wb}\hb
+3\,\frac{\pwwb}{\partial_u}h\,\partial_u\hb+3\,\partial_u h\, \frac{\pwwb}{\partial_u}\hb\bigg) \Bigg]
+ {\mathcal O}(\kappa^3).
\label{lagG}
\end{align}
The self-dual equation \eqref{SDGeom} is easily recovered from the general equations of motion by setting $\hb=0$, $h=-\partial_u^2 \phi/4$.

In momentum space, the equations of motion take the form
\begin{align}
h(k) =& \; \kappa\, \int \dd p_1  \dd p_2 \, \Big\{ 2 \, \frac{ k_u^2 }{p_{1u}^2p_{2u}^2} \, \frac{ X(p_1,p_2) \, F_{p_1 p_2}{}^k }{k^2} \, h(p_1) h(p_2) \nonumber \\ & \hspace{2.6cm} 
+  4 \, \frac{ p_{1u}^2 }{p_{2u}^2k_u^2} \, \frac{  \Xb(p_1,p_2) \,\Fb_{p_1 p_2}{}^k }{k^2} \, h(p_1) \hb(p_2) \Big\} 
\hspace{3cm}  \nonumber \\
& + \; \textrm{four-point vertex} \;+ \; {\mathcal O}(\kappa^3) .
\end{align}
We already see the squaring of the kinematic factors in the cubic vertices, when comparing with the analogous expression in Yang-Mills theory, equation \eqref{fullYMeomk}. More explicitly, we can determine the tree-level $++--$ (off-shell) amplitude. We saw in the Yang-Mills case \eqref{current} that the reference leg choice $p_{4u} \to 0$ was very useful. The same procedure leads now to
\begin{align}
{\mathcal M}&(1^+ 2^+ 3^- 4^-) = \kappa^2 \,
\varepsilon^+_{ww}(p_1)\varepsilon^+_{ww}(p_2)\varepsilon^-_{\wb\wb}(p_3)\varepsilon^-_{\wb\wb}(p_4) \,
\frac{16}{p_{4u}^2} \,
\frac{p_{3u}^2}{p_{1u}^2p_{2u}^2}\, \del(p_1+p_2+p_3+p_4) \nonumber \\
& \times \Bigg\{ \frac{ X(p_1,p_2)^2 \, \Xb(p_3,p_4)^2 }{{s}}
+ \frac{ X(p_1,p_2)^2 \, \Xb(p_3,p_4)^2 }{{u}}
+ \frac{ X(p_1,p_2)^2 \, \Xb(p_3,p_4)^2 }{{t}} 
\Bigg\} .
\label{currentG}
\end{align}
The $\Xb$ vertex associated with the reference leg is proportional to $X$, as we saw before. The amplitude is finite, due to the contribution from the polarization vectors, which are taken to be the squares of the Yang-Mills polarization vectors,
\begin{equation}
\varepsilon_{\mu\nu}^{\pm} (k) = \varepsilon_{\mu}^{\pm} (k) \, \varepsilon_{\nu}^{\pm} (k).
\end{equation}
Notice that it is crucial that the four-point vertex vanishes in the limit $p_{4u} \to 0$. This can be seen directly from the Lagrangian \eqref{lagG}: $1/\partial_u^{2}$ never acts on a single particle ($h$ or $\hb$) in the $\kappa^2$ contribution.

It is no surprise that the BCJ squaring \eqref{BCJgrav} holds for this MHV amplitude. In fact, under a certain assumption, we can show that it holds for any MHV amplitude. The argument is the same as the one used in Section~\ref{subsec:higherpoint} to show that the four-point vertex is irrelevant for MHV amplitudes, if we use the reference leg gauge. Again, we consider an $n$-point MHV amplitude $+\cdots+--$, and denote the number of internal lines by $I$, the number of $++-$ vertices by $n_+$, and the number of $+--$ vertices by $n_-$. The additional vertices are labeled by the number of incoming positive helicity lines $\sigma_+$ and the number of incoming negative helicity lines $\sigma_-$; the number of such vertices is denoted by $n_{\sigma_+\sigma_-}$. These additional vertices all satisfy $\sigma_+,\sigma_-\geq2$, which can be explicitly checked by extending the Lagrangian above to higher orders in $\kappa$. Then, the expressions analogous to (\ref{ngym}-\ref{nmym}) are
\begin{align}
n - 2 &= n_+ + n_- + \sum_{\sigma_+,\sigma_-} (\sigma_+ + \sigma_- -2) \,n_{\sigma_+\sigma_-}, \\
n-2 + I &= 2 n_+ + n_- + \sum_{\sigma_+,\sigma_-} \sigma_+ \,n_{\sigma_+\sigma_-}, \\
2 + I &= n_+ + 2 n_- + \sum_{\sigma_+,\sigma_-} \sigma_- \,n_{\sigma_+\sigma_-},
\end{align}
from which we conclude that
\begin{equation}
n_- + \sum_{\sigma_+,\sigma_-} (\sigma_- -1) \,n_{\sigma_+\sigma_-} =1.
\end{equation}
We assume that, as in Yang-Mills theory, the reference leg must be attached to a cubic vertex. We choose a negative helicity particle as our reference leg, so that $n_-=1$. It follows that the MHV amplitudes can be computed only from cubic vertices. For our assumption to be true, it is required that no higher-order term in the Lagrangian contains $1/\partial_u^{2}$ acting on a single particle. Ref.~\cite{Ananth:2008ik} presents the $\kappa^3$ contribution to the Lagrangian, and this is indeed the case. We conjecture that this is valid to all orders.


\section{Conclusions}
\label{sec:concl}

In the early sections of the paper, we described how a study of the self-dual sectors of Yang-Mills theory and of gravity sheds light on the BCJ relations. Cubic Feynman diagrams appear quite naturally in the self-dual sectors, and it is easy to see that the numerators of these diagrams have the properties which BCJ described. This occurs for a very simple reason: the numerators are built out of structure constants of an infinite dimensional Lie algebra. In the case of gravity, the presence of an infinite dimensional algebra seems entirely reasonable, and it is pleasing to see such an algebra play the role in gravity that the finite dimensional colour algebra plays in Yang-Mills theory. Indeed, Mason and Newman~\cite{Mason:1989ye} (see also \cite{Dunajski:2000iq}) previously proposed a very similar relationship between gauge theory and gravity. But it is quite remarkable that the KLT relations somehow work in reverse, so that this algebra is also present in the Yang-Mills theory. One of the main results of this paper is that the Yang-Mills cubic vertices are dressed by the straight product of the structure constants of the colour algebra and the kinematic algebra.

To make the kinematic algebra manifest in the case of MHV amplitudes of Yang-Mills theory, it was useful for us to work in the Chalmers-Siegel formulation of the theory. By choosing a special gauge, we removed the four-point vertex, so that the MHV amplitudes are computed off-shell by cubic diagrams. This is reminiscent of the BCFW computation of the MHV amplitudes. The close connection between BCFW and the Chalmers-Siegel action has already been pointed out by Vaman and Yao~\cite{Vaman:2010ez}. It seems that to clarify the structure of the kinematic algebra beyond MHV amplitudes will require understanding how to formulate these amplitudes in terms of cubic diagrams. Such a formulation must be related to the pure-spinor calculation of the BCJ numerators described by Mafra, Schlotterer and Stieberger \cite{Mafra:2011kj}. Similarly, it would be interesting to see the supersymmetric version of this algebra. The work of Broedel and Kallosh \cite{Broedel:2011ib} on the light-cone formulation of ${\mathcal N}=4$ SYM and ${\mathcal N}=8$ supergravity would be a good starting point.

On the gravitational side, we have seen how to compute MHV amplitudes from cubic diagrams in light-cone gauge, up to a mild assumption about the behaviour of the higher terms in the gravitational light-cone Lagrangian. The fact that the BCJ relations hold (at least on-shell) for all graviton amplitudes suggests that there is some cubic theory which can be used to compute these amplitudes. Finding this theory would be a great step forward in our understanding of both gauge theory and gravity.

\appendix

\acknowledgements

We thank Simon Badger, Rutger Boels, N. Emil Bjerrum-Bohr, Poul-Henrik Damgaard, Hidehiko Shimada and Thomas S\o ndergaard for useful discussions. DOC would like to thank the Science Institute of the University of Iceland for hospitality during the completion of this work. RM is supported by the Danish Council for Independent Research - Natural Sciences (FNU).


\end{document}